\documentclass[%
reprint,
superscriptaddress,
showpacs,
 amsmath,amssymb,
 aps,
]{revtex4-1}

\usepackage{graphicx}% Include figure files
\usepackage{dcolumn}% Align table columns on decimal point
\usepackage{bm}% bold math
\begin{document}

%\preprint{APS/123-QED}

\title{Supersymmetric Theory of Stochastic ABC Model: A Numerical Study}
\author{Igor V. Ovchinnikov}
\affiliation{Electrical Engineering Department, University of California at Los Angeles, Los Angeles, CA 90095.}
%\affiliation{APIC Corp., 5800 Uplander Way, Culver City, CA 90230}
\author{Yuquan Sun}
\affiliation{LMIB \& School of Mathematics and Systems Science, BeiHang University, Beijing 100191, China}
\author{Torsten A. En{\ss}lin}
\affiliation{Max-Planck-Institut f\"ur Astrophysik, Karl-Schwarzschildstr. 1, 85748 Garching, Germany}
\affiliation{Ludwig-Maximilians-Universit\"at M\"unchen, Geschwister-Scholl-Platz 1, 80539 Munich, Germany}
\affiliation{Technische Universit\"at M\"unchen, Exzellenzcluster Universe, Boltzmannstr. 2, 85748 Garching, Germany}
%\author{Raj Dutt}
%\affiliation{APIC Corp., 5800 Uplander Way, Culver City, CA 90230}
\author{Kang L. Wang}
\affiliation{Electrical Engineering Department, University of California at Los Angeles, Los Angeles, CA 90095.}

\date{\today}% It is always \today, today,

\begin{abstract}
In this paper, we investigate numerically the stochastic ABC model, a toy model in the theory of astrophysical kinematic dynamos, within the recently proposed supersymmetric theory of stochastics (STS). STS characterises stochastic differential equations (SDEs) by the spectrum of the stochastic evolution operator (SEO) on elements of the exterior algebra or differentials forms over the system's phase space, $X$. STS can thereby classify SDEs as chaotic or non-chaotic by identifying the phenomenon of stochastic chaos with the spontaneously broken topological supersymmetry that all SDEs possess. We demonstrate the following three properties of the SEO, deduced previously analytically and from physical arguments: the SEO spectra for zeroth and top degree forms never break topological supersymmetry, all SDEs possess pseudo-time-reversal symmetry, and each de Rahm cohomology class provides one supersymmetric eigenstate. Our results also suggests that the SEO spectra for forms of complementary degrees, \emph{i.e.}, $k$ and $\mathrm{dim} X -k$, may be isospectral.
\end{abstract}
\maketitle

%\tableofcontents

%%%%%%%%%%%%%%%%%%%%%%%%%%%%%%%%%%%%%%%%%%%%
%%%%%%%%%%%%%%%%%%%%%%%%%%%%%%%%%%%%%%%%%%%%
\section{Introduction}
%%%%%%%%%%%%%%%%%%%%%%%%%%%%%%%%%%%%%%%%%%%%
%%%%%%%%%%%%%%%%%%%%%%%%%%%%%%%%%%%%%%%%%%%%

The theory of stochastic differential equations (SDEs) has a long history and it provides many important insights on natural dynamics influenced by external noise (see, \emph{e.g.}, Refs.\cite{Kunita1,Baxendale1,Watanabe1,LaJen1,Oks10,Arn03} and Refs. therein). One of such insights has emerged recently as a result of the conjecture \cite{Mine0} that the theoretical essence of self-organized criticality \cite{Bak} may be the instanton-induced spontaneous breakdown of topological supersymmetry that all SDEs possess. Further work in this direction led to the formulation of the approximation-free supersymmetric theory of stochastics (STS) (see, \emph{e.g.}, Ref. \cite{Mine11} and Refs. therein).

From the mathematical point of view, the STS can be looked upon as a member of the cohomological or the Witten-type topological field theories \cite{TFT}, as a generalization of the Parisi-Sourlas quantization \cite{ParSour} from the Langevin SDEs to SDEs of any form, and as the application of the concept of the generalized transfer operator of the dynamical systems theory \cite{Rue02} to SDEs.

From the physical point of view, the importance of STS is in providing the theoretical picture for "dynamical long range order" (DLRO), known under such names as turbulence, chaos, self-organization, pattern formation, self-organized criticality, complex dynamics \emph{etc}. Within the STS, DRLO is the spontaneous breakdown of the topological or de Rahm supersymmetry. The existence of this supersymmetry in all SDEs is the algebraic representation of the phase-space continuity of the SDE-defined dynamics \cite{Mine11}. More specifically, two infinitely close points in the phase space will remain close during the SDE-defined evolution at any configuration of the stochastic noise. From this perspective, the spontaneous breakdown of topological supersymmetry can be interpreted as the breakdown of this property in the limit of the infinitely long evolution, represented, of course, by the non-supersymmetric ground state. In other words, in the limit of the infinitely long evolution, two close points in the phase space may not be close anymore and the system may be said to exhibit the butterfly effect and for this reason identified as chaotic. 

The supersymmetry breaking picture of chaotic behavior generalizes its classical picture from dynamical systems theory to stochastic dynamics \cite{Mine2}. Furthermore, this picture provides a rigorous explanation (via the Goldstone theorem) for the ubiquitous emergent long-range dynamical behavior in nature. Besides the butterfly effect, this emergent long-range dynamics also includes 1/f noise or the long-term memory effect and the Ritcher scale or the power-law statistics of sudden instantonic processes such as earthquakes, solar flares, neuroavalanches \emph{etc}.  

The centerpiece of the STS is the stochastic evolution operator (SEO) defined on the exterior algebra \cite{Nakahara} of the phase space, which is regarded as the Hilbert space of the stochastic model. The SEO describes the stochastically averaged SDE-induced actions on the wavefunctions, the elements of this Hilbert space. This stochastic averaging is possible since the Hilbert space as well as the SDE-induced actions on it are linear objects. In contrast, the operation of stochastic averaging on, say, SDE-defined trajectories cannot be defined in the general situation when the phase space is not a linear space. In this manner, the SEO permits to study interplay between stochastic and dynamical properties of the systems.

One of the important questions within the STS is the possible forms that the SEO spectrum can take, as these encode the system's characteristics of being chaotic, \emph{etc}. The key quantity here is the eigenvalue of the fastest growing eigenmode(s) of SEO, that must be identified as the ground state of the model. The problem of possible SEO spectra has been recently solved partly in Ref. \cite{Torsten} by establishing a connection between the STS and the theory of kinematic dynamo (KD, see,\emph{e.g.}, Refs. \cite{KD1,KD2,KD3,KDABC} and Refs. therein). The KD is the weak-magnetic-field limit of the more general astrophysical phenomenon of magnetic dynamos, \emph{i.e.}, the amplification of a magnetic field by a moving conducting medium (see, \emph{e.g.}, Refs. \cite{MD1,MD2,MD3,MD4,MD5,MD6} and Refs. therein). With the help of this STS-KD connection and using previous numerical results from the KD theory \cite{KD1,KD2,KD3,KDABC}, it has been established that SEOs with ground states with both real and complex eigenvalues are realizable.

The theory of KD deals, however, only with the non-supersymmetric eigenstates of the first and second degrees, the states that represent respectively the vector potential and the corresponding field tensor of the magnetic field. As such, the STS-KD connection cannot elucidate on other properties of the SEO. A few properties deduced previously analytically and from physical arguments remain to be proven or at least demonstrated numerically. Such a numerical demonstration is the goal of the present paper. 

Here, we numerically study the SEO of a stochastic ABC model, one of the toy models used in KD theory to mimic chaotic flows \cite{KDABC}. Our results support the following previously deduced properties of the SEO: \emph{(i)} the spectra of the top and the zeroth degrees do not break topological supersymmetry, \emph{(ii)} each de Rahm cohomology class provides one supersymmetric eigenstate, and \emph{(iii)} the overall spectra possess pseudo-time-reversal symmetry. In addition, our results suggest that the SEO with complementary degrees are isospectral -- a property of SEOs for which we do not have yet a theoretical explanation.

The paper is organized as follows. In Sec. \ref{sec:Nutshell}, we introduce the SEO for a general-form SDE with Gaussian white noise. In Sec. \ref{sec:Model}, we introduce the stochastic ABC model, whereas the details of the numerical realization of its SEO are given in the Appendix. In Sec. \ref{sec:Properties}, the properties of the SEO are discussed and exemplified by the numerical results for the stochastic ABC flow model. Sec. \ref{sec:Conclusion} concludes the paper.

%%%%%%%%%%%%%%%%%%%%%%%%%%%%%%%%%%%%%%%%%%%%
%%%%%%%%%%%%%%%%%%%%%%%%%%%%%%%%%%%%%%%%%%%%
\section{Stochastic Evolution with Gaussian White Noise} 
\label{sec:Nutshell}
%%%%%%%%%%%%%%%%%%%%%%%%%%%%%%%%%%%%%%%%%%%%
%%%%%%%%%%%%%%%%%%%%%%%%%%%%%%%%%%%%%%%%%%%%
The following general-form SDE is of our interest:
\begin{eqnarray}
\dot {\bm x}(t) = {\bm F}({\bm x}(t)) + (2\Theta)^{1/2}{\bm e}_a({\bm x}(t))\xi^a(t)\equiv {\mathcal {\bm F}}(t). \label{SDE}
\end{eqnarray}
Here and in the following the summation is assumed over the repeated indices; $\bm x\in X$ is a point in the phase space, $X$, which is assumed to be a topological manifold; $\bm F({\bm x})\in TX_{\bm x}$ is a flow vector field on $X$ at $\bm x$; ${\bm e}_a({\bm x})\in TX_{\bm x}, a=1,...$ is a set of vector fields on $X$ with $a$ being the parameter running over these vector fields; $\xi^a(t)$ is a set of the Gaussian white noise variables.

For each configuration of the noise, Eq.(\ref{SDE}) defines the family of trajectories in $X$. Alternatively, this family of trajectories can be looked upon as a two-parameter family of diffeomorphisms of $X$ on itself \cite{Mine11}, with the two parameters being the initial, $t$, and the final, $t'>t$, time moments of temporal evolution,
\begin{eqnarray}
M_{t't}:X\to X.\label{Maps}
\end{eqnarray}
These diffeomorphisms induce actions or pullbacks on differential forms,
\begin{eqnarray}
\hat M^*_{tt'}:\Omega^k&\to&\Omega^k, \label{Actions}\\
\psi^{(k)}({\bm x}) &=& \psi^{(k)}_{i_1...i_k}({\bm x})dx^{i_1}\wedge...\wedge dx^{i_k}\in \Omega^k({\bm x}), \label{k_forms}\\
(\hat M^*_{tt'}\psi^{(k)})({\bm x}) &=& \tilde \psi^{(k)}_{\tilde i_1...\tilde i_k}({\bm x})dx^{\tilde i_1}\wedge...\wedge dx^{\tilde i_k}, \nonumber\\
\tilde \psi^{(k)}_{\tilde i_1...\tilde i_k}({\bm x}) &=&\psi^{(k)}_{i_1... i_k}(M_{tt'}({\bm x}))TM^{i_1}_{\tilde i_1}({\bm x})...TM^{i_k}_{\tilde i_k}({\bm x}), \nonumber\\
TM^{i}_{j}({\bm x})&=& \frac{\partial M_{tt'}^{i}({\bm x})}{\partial {\bm x}^j}.\label{TangentMap}
\end{eqnarray}
Here, $\Omega^k$ denotes the space of all differential forms of degree $k$, or $k$-forms, $\psi^{(k)}$, defined in Eq.(\ref{k_forms}) via the contravariant antisymmetric tensor, $\psi^{(k)}_{i_1...i_k}$, and the wedge or antisymmetric product of differentials, $\wedge$. Eq.(\ref{TangentMap}) is known as the tangent map. 

The action in Eq.(\ref{Actions}) is the most natural construction from the mathematical point of view. It can be looked upon as the formal change of variables in a differential form induced by maps that are inverse to the forward maps in Eq.(\ref{Maps}). The fact that these maps are inverse is reflected in the reversed order of $t'$ and $t$ in Eq.(\ref{Actions}) as compared to that in Eq.(\ref{Maps}). The reason for this seeming reversion of time is discussed in detail in Sec.2.1 of Ref.\cite{Mine11}. 

The entire exterior algebra of $X$, \emph{i.e.}, the space of the differential forms of all degrees,
\begin{eqnarray}
\Omega(X) = \bigoplus_{k=0}^{D}\Omega^k(X),
\end{eqnarray}
with $D=\mathrm{dim} X$, is the Hilbert space of the model. The top differential forms of maximal degree $D$ can be interpreted as the total probability distributions, whereas differential forms of lesser degrees can be looked upon, at least locally on $X$, as the conditional probability distributions \cite{MineChaos}.

The infinitesimal action of the SDE-defined diffeomorphisms can be given via the stochastic flow equation (SFE),
\begin{eqnarray}
\partial_t \psi(t)  &=& (\Delta t)^{-1}\left(M^*_{(t-\Delta t)t}-\hat 1_{\Omega}\right)\psi(t) \nonumber\\
&=&-\hat {\mathcal L}_{{\mathcal {\bm F}}(t)}\psi(t) \nonumber \\&=& -\left(\hat {\mathcal L}_{\bm F} + (2\Theta)^{1/2}\xi^a(t)\hat {\mathcal L}_{{\bm e}_a}\right)\psi(t),
\end{eqnarray}
where $\hat {\mathcal L}_{{\mathcal {\bm F}}(t)}$ denotes the Lie of physical derivative along ${{\mathcal {\bm F}}(t)}$. The SFE follows immediately from the understanding of the Lie derivative as of the infinitesimal pullback of the SDE-defined flow. Accordingly, the finite-time pullback is given as,
\begin{eqnarray}
\hat M^{*}_{tt'} &=& {\mathcal T} e^{-\int_t^{t'}d\tau \hat {\mathcal L}_{{\mathcal F}(\tau)}}=\hat 1_\Omega - \int_t^{t'}d\tau_1\hat {\mathcal L}_{{\mathcal F}(\tau_1)} \nonumber\\&+& \int_t^{t'}d\tau_1\hat {\mathcal L}_{{\mathcal F}(\tau_1)}\int_t^{\tau_1}d\tau_2\hat {\mathcal L}_{{\mathcal F}(\tau_2)}...\label{Chron}
\end{eqnarray}
Here $\mathcal T$ denotes the operation of chronological ordering, which is necessary as in the general case $\hat {\mathcal L}_{{\mathcal {\bm F}}(t)}$'s at different time moments of the evolution do not commute.

The pullback is a linear operator on a linear Hilbert space. Thus, this operator can be averaged over the configurations of the noise. Such stochastic averaging leads to the finite-time stochastic evolution operator (SEO),
\begin{eqnarray}
\hat{\mathcal M}_{t't} = \left< M^{*}_{tt'}\right>_\text{Ns},
\end{eqnarray}
where brackets denote the stochastic averaging. 

For the white noise case only, the finite-time SEO can be expressed via the (infinitesimal) SEO, $\hat H$, in the following manner,
\begin{eqnarray}
\hat{\mathcal M}_{t't} = e^{-(t'-t)\hat H},
\end{eqnarray}
so that the infinitesemal evolution of the wavefunctions is given by the following stochastic evolution equation,
\begin{eqnarray}
\partial_t\psi(t) = - \hat H \psi(t).
\end{eqnarray}
The explicit form of $\hat H$ can be readily established using the following definition,
\begin{eqnarray}
\hat H = (\Delta t)^{-1}\left<\hat 1_\Omega - M^*_{(t-\Delta t)t}\right>_\text{Ns}. 
\end{eqnarray}
With the help of Eq.(\ref{Chron}) and the standard expectation values of the Gaussian white noise variables, $\langle \xi^a(\tau_1)\rangle_\text{Ns} = 0$ and $\langle \xi^{a_1}(\tau_1)\xi^{a_2}(\tau_2)\rangle_\text{Ns} = \delta^{a_1a_21}\delta(\tau_1-\tau_2)$, one finds that
\begin{eqnarray}
\hat H = \hat {\mathcal L}_{{\bm F}} - \Theta \hat {\mathcal L}_{{\bm e}_a}\hat {\mathcal L}_{{\bm e}_a}.\label{SEO}
\end{eqnarray}
Our next goal is to discuss the properties of the SEO and exemplify these properties using our numerical results for the stochastic ABC model that we discuss in the next section.
\begin{figure*}[t]
    \centering
    \includegraphics[width=18cm,height=6.3cm]{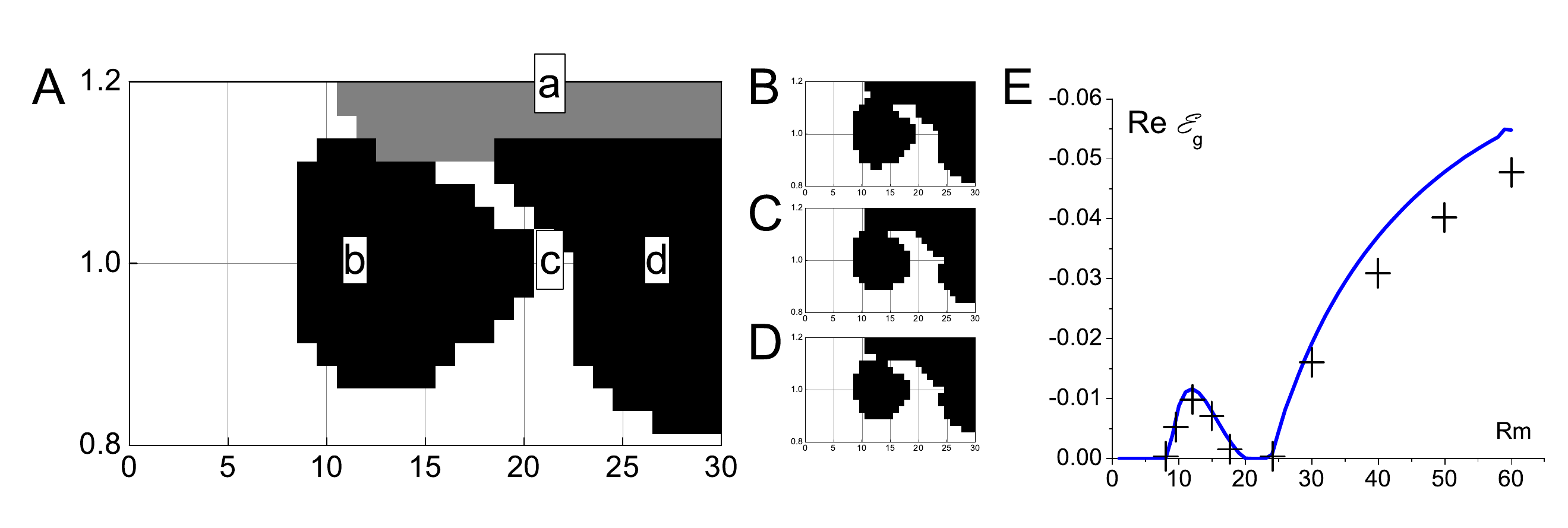}
    \caption{(color online) ({\bf A}) The phase diagram of the stochastic ABC model for $A=B=1$, $0.8<C<1.2$, $1<\mathrm{Rm}<30$, and the grid parameter $N=30$. The black and grey areas are the regions where the supersymmetry is spontaneously broken by non-supersymmetric ground states with, respectively, complex and real eigenvalues. Lower case letters $(a,b,c,d)$ indicate the points, the full spectra of which are presented in Fig. \ref{Figure_2}. ({\bf B-D}) The same for grid parameters $N=35, 40$, and $45$, and without the separation into the subregions with the real and complex ground state eigenvalues. ({\bf E}) The real part of the ground state's eigenvalue, $Re \mathcal{E}_g$, for $A=B=C=1$ and $1<\mathrm{Rm}<60$. Crosses represent the results from Fig. 1 of Ref. \cite{KDABC} read off by eye.
}
    \label{Figure_1}
\end{figure*}

%%%%%%%%%%%%%%%%%%%%%%%%%%%%%%%%%%%%%%%%%%%%%%%%%%%%
%%%%%%%%%%%%%%%%%%%%%%%%%%%%%%%%%%%%%%%%%%%%%%%%%%%%
\section{Stochastic ABC model}
\label{sec:Model}
%%%%%%%%%%%%%%%%%%%%%%%%%%%%%%%%%%%%%%%%%%%%%%%%%%%%
%%%%%%%%%%%%%%%%%%%%%%%%%%%%%%%%%%%%%%%%%%%%%%%%%%%%

The stochastic ABC model on a 3-torus is defined by its flow vector field and the three "Euclidian" ${\bm e}$'s,
\begin{eqnarray}
{\bm F}_{ABC} = A\cdot(\sin z,\cos z,0)^\mathrm{T} + B\cdot(0,\sin x, \cos x)^\mathrm{T}\nonumber\\ + C\cdot (\cos y, 0, \sin y)^\mathrm{T},\nonumber\\
{\bm e}_1 = (1,0,0)^\mathrm{T}, {\bm e}_2 = (0,1,0)^\mathrm{T}, {\bm e}_3 = (0,0,1)^\mathrm{T}.\label{ABC_e_s}\label{ABC}
\end{eqnarray}
The vector fields $e$'s represent the flat metric on the 3-torus, for which the diffusion Laplacian is the Hodge Laplacian,
\begin{eqnarray}
\hat {\mathcal L}_{{\bm e}_a}\hat {\mathcal L}_{{\bm e}_a} = - [\hat d, \hat d^\dagger],\label{HodgeLap}
\end{eqnarray}
where $\hat d^\dagger = - \hat \imath_j \delta^{ij}\partial/\partial x^i$ is the so-called co-differential operator (see Eq.(\ref{InterriorMultiplication}) below for the definition of $\hat \imath_j$). 

Eqs.(\ref{SEO}), (\ref{ABC}), and (\ref{HodgeLap}) provide the SEO of the stochastic ABC model,
\begin{eqnarray}
\hat H_{ABC} = \hat {\mathcal L}_{{\bm F}_{ABC}} + \mathrm{Rm}^{-1} [\hat d, \hat d^\dagger].
\end{eqnarray}
where the inverse temperature, $\mathrm{Rm}=\Theta^{-1}$, is known in the KD theory as the magnetic Reynolds number and it parametrises physical diffusivity of the magnetic field in units of the turbulent diffusivity of the flow. If $\Theta=0$, the magnetic field is perfectly frozen into the flow of the conducting medium.

The procedure for the construction of the SEO on the square lattice of the 3-torus is described in the Appendix. There, one of the parameters of the model is the lattice constant, $a=2\pi/N$, where $N$ is the number of lattice sites in each of the three directions. It is understood that the finer the lattice, or equivalently the larger the $N$, the more accurate numerical representations of the SEO one obtains and thus the more trustworthy results one gets. On the other hand, the computer time taken up by the diagonalization procedure involved in the spectrum calculation increases dramatically with $N$. In order to find balance between accuracy and the availability of the machine-time resources, we compared the results for different $N$'s. 

In Fig. (\ref{Figure_1}), the phase diagrams of the stochastic ABC model for the four different choices of $N$ are presented. As is seen, the phase diagrams for $N=30,35,40,$ and $N=45$ are qualitatively the same. From this observation we conclude that for this particular range of parameters ($0.8<C<1.2$ and $1<\mathrm{Rm}<30$), $N=30$ already provides a reasonably good approximation for the SEO. All the subsequent numerical results are therefore obtained for this particular choice of $N=30$. 

To improve our confidence in the results with $N=30$, we compared the value of the real part of the ground state of the model with $A=B=C=1$ with the results obtained in Ref. \cite{KDABC}. As can be seen in Fig. \ref{Figure_1}E, our method with $N=30$ reproduces the results of Ref. \cite{KDABC} qualitatively well. Note that only the situations with spontaneously broken supersymmetry can be compared ($8<\mathrm{Rm}<21$ and $\mathrm{Rm}>23$). The point is that the employed method of finding the "fastest growing" eigenvalue provides zero eigenvalue of the supersymmetric states for the situations with unbroken supersymmetry. In Ref. \cite{KDABC}, on the other hand, only the $\hat d$-exact non-supersymmetric eigenstates that represent the field tensor of the magnetic field of the KD effect are considered. This is the reason why the real part of the ground state eigenvalue of our method coincides with the magnetic field growth rate in Ref.\cite{KDABC} only when the supersymmetry is spontaneously broken and the ground state is the $\hat d$-exact non-supersymmetric eigenstate of the second degree with the fastest growth rate.

\begin{figure*}[t]
    \centering
    \includegraphics[width=18.0cm,height=10.0cm]{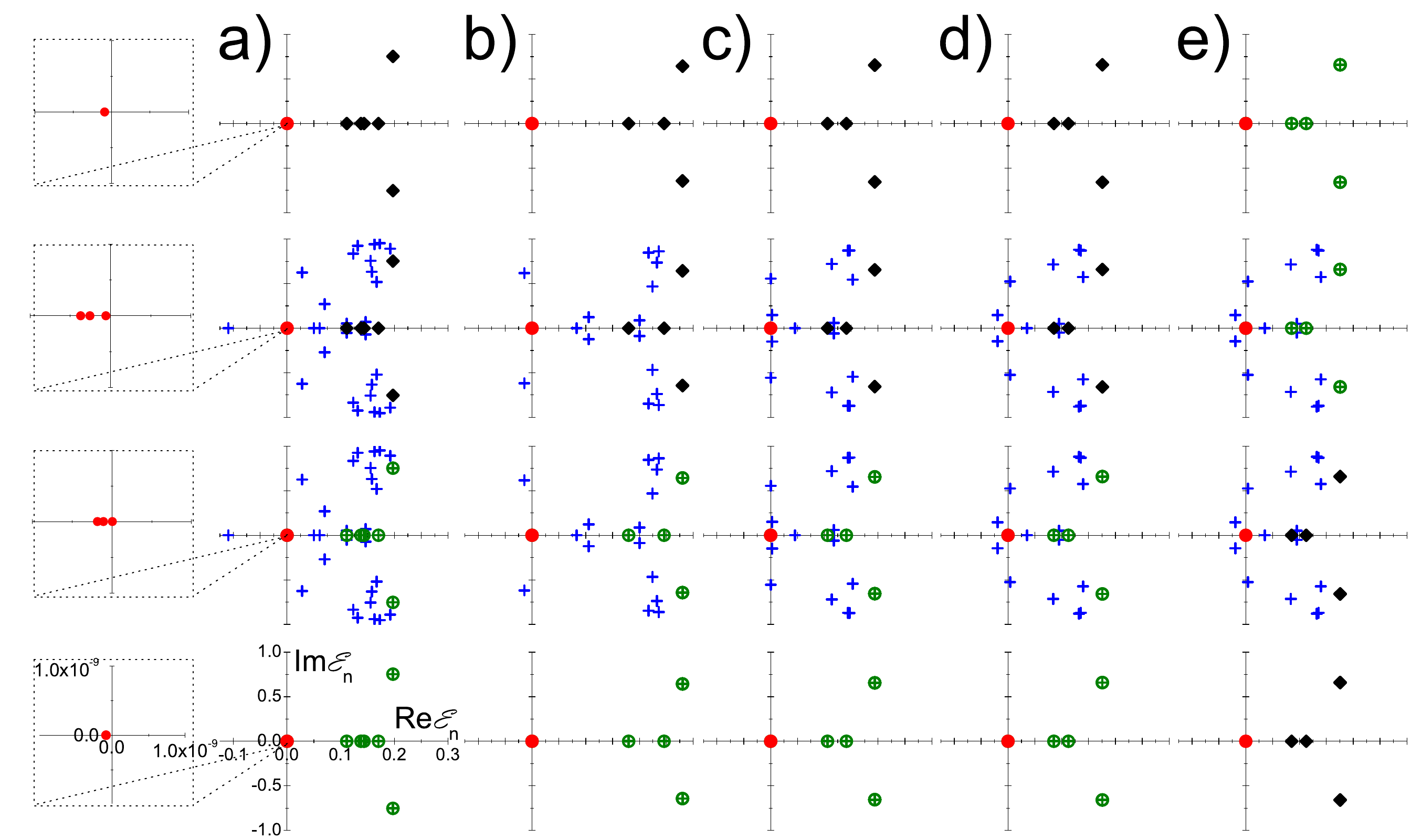}
    \caption{(color online) Spectra of SEO at different parameters of the model. The grid parameter $N=30$. The green circled crosses, blue crosses, and black diamonds represent the boson-fermion pairs of non-supersymmetric eigenstates of, respectively, degrees 0 and 1, 1 and 2, and 2 and 3. Red circles at the origin represent supersymmetric states. ({\bf a}) $A=B=1$, $C=1.2$, $\mathrm{Rm}=22$. The degrees are in the incremental order ($k=0,1,2,3$) from bottom to top. The scale is the same everywhere and is given in the $k=0$ (bottom) spectrum. ({\bf Insets}) Supersymmetric states have zero eigenvalues well within the numerical error of calculations. The scale is the same everywhere and is given in the $k=0$ (bottom) inset. ({\bf b})-({\bf d}) SEO spectra for $A=B=C=1$ and, respectively, $\mathrm{Rm}=15,22$, and $28$. ({\bf e}) The SEO spectra for $A=B=C=-1$ and $\mathrm{Rm}=28$. The order of the degrees is reversed ($k=3,2,1,0$ from bottom to top) for the comparison with ({\bf d}) as discussed in the text.}
    \label{Figure_2}
\end{figure*}

%%%%%%%%%%%%%%%%%%%%%%%%%%%%%%%%%%%%%%%%%%%%%%%%%%%%
%%%%%%%%%%%%%%%%%%%%%%%%%%%%%%%%%%%%%%%%%%%%%%%%%%%%
\section{Properties of the SEO}
\label{sec:Properties}
%%%%%%%%%%%%%%%%%%%%%%%%%%%%%%%%%%%%%%%%%%%%%%%%%%%%
%%%%%%%%%%%%%%%%%%%%%%%%%%%%%%%%%%%%%%%%%%%%%%%%%%%%

The goal of this section is to discuss the properties of the SEO and demonstrate them using the numerical results obtained for the stochastic ABC model. 

The SEO is a real operator and consequently its eigenstates are either real or come in complex conjugate pairs known in the dynamical systems theory as Ruelle-Pollicott resonances. Also, for non-degenerate noise-induced metric on $X$, $e_a^i({\bm x}) e_a^j({\bm x})$, which is the case for the stochastic ABC model, and for non-zero temperature and/or reciprocal of the Reynolds number, the SEO is elliptic and the real part of its eigenvalues is bounded from below. Both of these properties can be observed in Fig.~\ref{Figure_2}, where the spectra of the SEO for different parameters of the stochastic ABC model are presented.

An operator with these properties is pseudo-Hermitian  \cite{Mostafa}. It has the so-called complete bi-orthogonal eigensystem. Furthermore, the SEO does not "mix" differential forms of different degrees, \emph{i.e.}, it conserves the number of fermions (see below), so that it can be looked upon as a block diagonal operator,
\begin{eqnarray}
\hat H = \mathrm{diag} (\hat H^{(D)}...\hat H^{(0)}), \hat H^{(k)}:\Omega^k\to\Omega^k,
\end{eqnarray}
with each $\hat H^{(k)}$ being pseudo-Hermitian. The eigensystem of the SEO can now be introduced,
\begin{eqnarray}
\langle n_k | \hat H^{(k)} &=& \langle n_k | {\mathcal E}_{n_k},\\
\hat H^{(k)} |n_k\rangle &=& {\mathcal E}_{n_k} |n_k\rangle,\\
\langle n_k|m_k\rangle &=& \delta_{n_km_k}.
\end{eqnarray}
Here, kets of the eigenstates are k-forms, $|n_k\rangle \equiv \psi_{n_k}\in\Omega^k$, and the bras are the differential forms of the complementary degrees, $\langle n_k | \equiv \bar \psi_{n_k}\in\Omega^{D-k}$, so that the overlap $\langle n_k|n_k\rangle \equiv \int_X \bar \psi_{n_k}\wedge \psi_{n_k}$ does not vanish.

To establish the supersymmetric structure underlying stochastic evolution let us introduce fermionic or Grassmann anticommuting variables, $\chi^i\equiv dx^i\wedge$, $\chi^i\chi^j \equiv dx^i\wedge dx^j = - dx^j\wedge dx^i =- \chi^j \chi^i,$ \emph{etc}. In these new notations, the wavefunction (\ref{k_forms}) can be given as,
\begin{eqnarray}
\psi^{(k)}({\bm x}) = \psi^{(k)}_{i_1...i_k}({\bm x})\chi^{i_1}...\chi^{i_k}.
\end{eqnarray}
Let us also recall the Cartan formula for the Lie derivative,
\begin{eqnarray}
\hat{\mathcal L}_{\bm F} = [\hat d, {\bm F}^i\hat{\imath}_i],\label{CartanFormula}
\end{eqnarray}
where,
\begin{eqnarray}
\hat d:\Omega^{k}\to\Omega^{k+1}, \hat d = \chi^i\frac{\partial }{\partial x^i},  \label{susy}
\end{eqnarray}
is the exterior derivative or de Rahm operator, and
\begin{eqnarray}
\hat \imath_i:\Omega^{k}\to\Omega^{k-1}, \hat \imath_i = \frac{\partial}{\partial \chi^i}, \label{InterriorMultiplication}
\end{eqnarray}
is the interior multiplication operator. The commutator in Eq.(\ref{CartanFormula}) denotes bi-graded commutator, which is an anticommutator if both operators are "bosonic", \emph{i.e.}, have odd number of fermionic operators, and it is a commutator otherwise. In particular, the bi-graded commutator in Eq.(\ref{CartanFormula}) is an anticommutator.

With the help of the nilpotency property of the exterior derivative,
\begin{eqnarray}
[\hat d[\hat d, \cdot]]=0,\label{Nilpotent}
\end{eqnarray}
and knowing that the commutator with the exterior derivative is a bi-graded differentiation,
\begin{eqnarray}
[\hat d, \hat A \hat B] = [\hat d, \hat A] \hat B + (-1)^{\mathrm{deg} A}\hat A[\hat d, \hat B],
\end{eqnarray}
with $\mathrm{deg} A$ being the degree of an operator defined as the number of $\chi$'s minus the number of $\partial/\partial\chi$'s, one readily finds that,
\begin{eqnarray}
\hat H  = [\hat d, \hat {\bar d}], 
\end{eqnarray}
where,
\begin{eqnarray}
\hat {\bar d} = {\bm F}^i\hat{\imath}_i - \Theta {\bm e}_a^i \hat{\imath}_i \hat{\mathcal L}_{{\bm e}_a}.
\end{eqnarray}
Using now the nilpotency property in Eq.(\ref{Nilpotent}), one finds that the SEO is commutative with the exterior derivative,
\begin{eqnarray}
[\hat d, \hat H ] = 0.
\end{eqnarray}
In other words, $\hat d$ is a symmetry of the SEO or rather a supersymmetry because, as it is seen from Eq.(\ref{susy}), it kills a commuting or bosonic variable and substitutes it with an anti-commuting or fermionic variable.

In physics, symmetries reveal themselves as protected degeneracies of the eigenstates of evolution operators. More technically, it is said that the multiplets, \emph{i.e.}, the eigenstates corresponding to a degenerate eigenvalue, are irreducible representations of the corresponding symmetry group. In case of the topological supersymmetry, $\hat d$, the multiplets are the non-supersymmetric doublets or the boson-fermion pairs, \emph{i.e.}, all non-supersymmetric eigenstates come in pairs of even and odd degrees, $|\vartheta\rangle$ and $\hat d|\vartheta\rangle$. It can be shown that all eigenstates with non-zero eigenvalues are non-supersymmetric as is also seen in Fig.\ref{Figure_2}, where the non-supersymmetric pairs of eigenstates are indicated using the same symbol (circled crosses, crosses, and diamonds).

Some of the eigenstates are supersymmetric singlets such that $\hat d|\theta\rangle=0$ and no state $|\theta'\rangle$ exists such that $|\theta\rangle = \hat d|\theta'\rangle$. In fact, this property of the supersymmetric eigenstates is nothing else but the requirement for a state to be non-trivial in de Rahm cohomology \cite{Nakahara}. 

All supersymmetric states have strictly zero eigenvalue and each de Rahm cohomology class provides one supersymmetric eigenstate, because otherwise the eigensystem of the SEO would not be complete, which would be in contradiction with the idea that elliptic pseudo-Hermitian operators on compact phase spaces have complete eigensystems \cite{Mostafa}.

The fact that each de Rahm cohomology class provides one zero-eigenvalue supersymmetric eigenstate can be observed in Fig. \ref{Figure_2}a, where insets zoom into the small area around the origin. It is seen that the number of eigenstates with zero eigenvalue (within the numerical accuracy of our calculations) equals the Betti number of the same degree, \emph{i.e.}, the number of different de Rahm cohomology classes of a given degree \cite{Nakahara}, which in the case of 3-torus is $1,3,3,1$ for $k=0,1,2,3$, respectively.

The ground states of the model are the ones with the fastest "growth rate" according to their temporal evolution $\sim e^{-\mathcal E_k t}$, \emph{i.e.}, the states with the minimal real part of their eigenvalues. When the zero-eigenvalue supersymmetric states are the ground states, it is said that the supersymmetry is unbroken. Among the five spectra presented in Fig.\ref{Figure_2}, only $c)$ has unbroken supersymmetry. For all the other spectra, the ground states have non-zero eigenvalues, real for $a)$ and complex for $b),d)$, and $e)$. For these spectra, the topological supersymmetry is broken spontaneously because the ground states are non-supersymmetric. The phenomenon of the spontaneous topological supersymmetry breaking can be looked upon as the stochastic generalization of deterministic chaos, \cite{Mine2} whereas in the KD theory, it corresponds to the existence of the exponentially growing modes of the magnetic field  \cite{Torsten}.

It can be shown that all models possess pseudo-time reversal symmetry \cite{Mine11}. This is the symmetry of the SEOs of two SDEs related to each other by the reversal of time,
\begin{eqnarray}
\hat H = \hat S^{-1} \hat H_{\mathfrak T} \hat S.\label{Similar}
\end{eqnarray}
Here $\hat H_{\mathfrak T}$ is the SEO of the SDE with the reversed flow, $F\to-F$, and reversed noise vector fields, $e\to -e$ (the reversing of $e$'s has no effect in Eq.(\ref{SEO}), however), and $\hat S:\Omega^{k}\to\Omega^{D-k}$ is some invertible operator. Any two operators related by a similarity transformation such as Eq.(\ref{Similar}) are isospectral. Thus, $\mathrm{spec}\hat H^{(k)}=\mathrm{spec}\hat H^{(D-k)}_{\mathfrak T}$. This symmetry of SEO is demonstrated in Fig. \ref{Figure_2}d) and e), where the spectra of $\hat H$ and $\hat H_{\mathfrak T}$ are given for $A=B=C$ and $\mathrm{Rm}=28$. This isospectrality will be used again in a moment to conclude that $\hat H^{(0)}$ alone never breaks supersymmetry spontaneously, just as $\hat H^{(D)}$.

As already mentioned, the wavefunctions of the top (or $k=3$) degree represent total probability distributions. Also, all non-supersymmetric eigenstates of the top degree are $\hat d$-exact, \emph{i.e.}, of the form $\hat d|\vartheta\rangle$, $\vartheta\in\Omega^{D-1}$. This suggests that the integral over such eigenstate over $X$ is zero and, consequently, somewhere on $X$ these non-supersymmetric eigenstates are negative. This leads to the conclusion that the supersymmetry cannot be broken spontaneously by the SEO of the top degree ($\hat H^{(D)}$). Indeed, if it is a non-supersymmetric eigenstate(s) that has the fastest growing rate in $\Omega^D$, then an arbitrary total probability distribution would become negative somewhere on $X$ after a sufficiently long temporal evolution when this non-supersymmetric ground state would provide a dominant contribution. Negative total probability distributions, on the other hand, are unphysical. 

Also, since $\hat H_{\mathfrak T}^{(D)}$ never breaks topological supersymmetry and, at the same time, it is isospectral to $\hat H^{(0)}$, we conclude that $\hat H^{(0)}$ also never breaks topological supersymmetry spontaneously. The fact that neither $\hat H^{(D)}$ nor $\hat H^{(0)}$ break topological supersymmetry spontaneously can be observed in all the spectra presented in Fig. \ref{Figure_2}.

At last, our results presented in Fig. \ref{Figure_2} as well as those obtained but not presented in this paper seemingly suggest that the SEO of complementary degrees are isospectral, $\mathrm{spec} \hat H^{(k)}=\mathrm{spec} \hat H^{(D-k)}$. As of this moment, we do know under what conditions this symmetry of the SEO is present and what are its mathematical and/or physical origins.

%%%%%%%%%%%%%%%%%%%%%%%%%%%%%%%%%%%%%%%%%%%%%%%%
%%%%%%%%%%%%%%%%%%%%%%%%%%%%%%%%%%%%%%%%%%%%%%%%
\section{Conclusion}
\label{sec:Conclusion}
%%%%%%%%%%%%%%%%%%%%%%%%%%%%%%%%%%%%%%%%%%%%%%%%
%%%%%%%%%%%%%%%%%%%%%%%%%%%%%%%%%%%%%%%%%%%%%%%%
In this paper, we numerically investigated the stochastic evolution operator of stochastic ABC model. The following general properties of the stochastic evolution operators, that were known previously, are confirmed: every de Rahm cohomology class provides one supersymmetric eigenstate; the stochastic evolution operators of the zeroth and top degrees separately never break topological supersymmetry; and stochastic models possess pseudo-time-reversal symmetry. In addition, our results suggest that the stochastic evolution operators of complementary degrees may be isospectral. We hope that further work on the STS will reveal the mathematical and/or physical reasons that stand behind of this symmetry of the stochastic evolution operators.  

\acknowledgments
KLW would like to acknowledge the support from Raytheon endowed professorship. YQS would like to thank National Science Foundation of China for support (Grant No. 11201020) and Device Research Laboratory of the Electrical Engineering Department of UCLA for hospitality during his visit in 2015-2016. IVO and TAE would like to acknowledge the partial support of this work from Excellence Cluster Universe.

\begin{figure*}
    \centering
    \includegraphics[width=16cm,height=5.8cm]{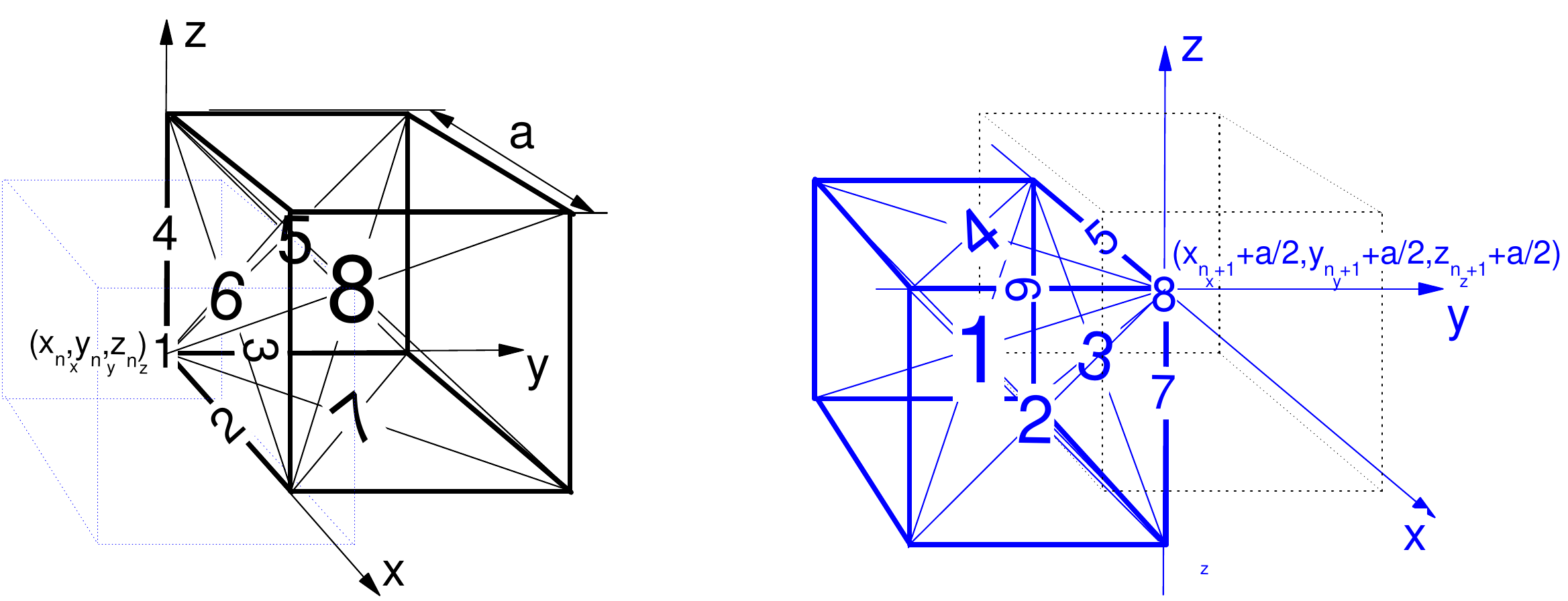}
    \caption{(color online) {\bf (left)} The square lattice partitions the phase space into a union of 3D cubes. The collection of all the cubes, their faces, edges, and points/vortexes constitute a cubic CW complex of the phase space. The basis of the Hilbert space are the Poincar\'e duals of the elements of this cubic CW complex. The indexes of the kets $(1,...,8)$, indicated explicitly, correspond to those in Eq.(\ref{Kets}). The dotted cube of thin lines represent the adjacent elementary cell of the dual lattice. $a$ is the lattice constant. {\bf (right)} The Poincar\'e duals of the cubic CW complex of the dual lattice is the basis of the dual Hilbert space given in Eq.(\ref{Bras}).}
    \label{Figure_A1}
\end{figure*}

\section{Appendix: Explicit Construction of the SEO on the Lattice.}
The goal in this Appendix is to discuss the construction of the numerical version of the SEO. First, we define the square lattice,
\begin{eqnarray*}
    {\bm x}_{{\bm n}} = (x_{n_x},y_{n_y},z_{n_z}),  
    {\bm n} = (n_x,n_y,n_z),
\end{eqnarray*}
where $x_{n_x} = 2 \pi (n_x-1)/N$ and similar for $y$'s and $z$'s, and indexes $n_{x,y,z}= 1,...,N$ run over the grid in the corresponding dimensions. The phase space is a 3-torus so that $x_{N+1}\equiv x_1$ and the same for other dimensions.

The collection of the elementary cubic cells, their faces, edges, and points/vertexes constitute the so-called cubic CW complex. The basis of the Hilbert space representing the lattice version of the exterior algebra is the so-called Poincar\'e duals of the elements of this cubic CW complex (see Fig.\ref{Figure_A1}). The Poincar\'e duals of submanifolds, in their turn, are the differential forms that are constant functions (with no differentials) on these submanifolds and that are delta-functional distributions (with differentials) in the transverse directions. Accordingly,
\begin{subequations}
\label{Kets}
\begin{eqnarray}
|1, {\bm n}\rangle &=& \Delta_{n_x}(x)\chi^x \Delta_{n_y}(y)\chi^y\Delta_{n_z}(z)\chi^z,\\
|2, {\bm n}\rangle &=& \theta_{n_x}(x)\Delta_{n_y}(y)\chi^y\Delta_{n_z}(z)\chi^z\\
|3, {\bm n}\rangle &=& - \Delta_{n_x}(x)\chi^x \theta_{n_y}(y)\Delta_{n_z}(z)\chi^z,\\
|4, {\bm n}\rangle &=& \Delta_{n_x}(x)\chi^x\Delta_{n_y}(y)\theta_{n_y}(y)\chi^y,\\
|5, {\bm n}\rangle &=& \Delta_{n_x}(x)\chi^x\theta_{n_y}(y)\theta_{n_z}(z),\\
|6, {\bm n}\rangle &=& \theta_{n_x}(x)\Delta_{n_y}(y)\chi^y\theta_{n_z}(z),\\
|7, {\bm n}\rangle &=& 
\theta_{n_x}(x)\theta_{n_y}(y)\Delta_{n_z}(z)\chi^z,\\
|8, {\bm n}\rangle &=& 
\theta_{n_x}(x)\theta_{n_y}(y)\theta_{n_z}(z),
\end{eqnarray}   
\end{subequations}
where we use fermionic notations for differentials, \emph{e.g.}, $\chi^x=dx\wedge$, and we introduced functions
\begin{eqnarray}
\Delta_{n_x}(x) = \delta(x-x_{n_x}),\label{delta_ket}
\end{eqnarray}
and 
\begin{eqnarray}
\theta_{n_x}(x) = a^{-1}\left\{\begin{array}{rcl} 1&,&x_{ n_x}<x<x_{n_x+1},\\0&,& \text{otherwise}\end{array}\right.,\label{theta_ket}
\end{eqnarray}
with $a = 2\pi/N$ being the lattice constant. Functions $\Delta$ and $\theta$ are defined similarly for the other two dimensions.

The basis of the dual Hilbert space is the Poincar\'e duals of the dual lattice (see Fig.\ref{Figure_A1}): 
\begin{subequations}
\label{Bras}
\begin{eqnarray}
\langle 1, {\bm n}| &=& 
\tilde\theta_{n_x}(x)\tilde\theta_{n_y}(y)\tilde\theta_{n_z}(z),\\
\langle 2, {\bm n}| &=& \tilde\Delta_{n_x}(x)\chi^x \tilde\theta_{n_y}(y)\tilde\theta_{n_z}(z),\\
\langle 3, {\bm n}| &=& \tilde\theta_{n_x}(x)\tilde\Delta_{n_y}(y)\chi^y\tilde\theta_{n_z}(z),\\
\langle 4, {\bm n}| &=& 
\tilde\theta_{n_x}(x)\tilde\theta_{n_y}(y)\tilde\Delta_{n_z}(z)\chi^z,\\
\langle 5, {\bm n}| &=& \tilde\theta_{n_x}(x)\tilde\Delta_{n_y}(y)\chi^y\tilde\Delta_{n_z}(z)\chi^z,\\
\langle 6, {\bm n}| &=& -\tilde\Delta_{n_x}(x)\chi^x \tilde\theta_{n_y}(y)\tilde\Delta_{n_z}(z)\chi^z,\\
\langle 7, {\bm n}| &=& \tilde\Delta_{n_x}(x)\chi^x\tilde\Delta_{n_y}(y)\chi^y\tilde\theta_{n_z}(z),\\
\langle 8, {\bm n}| &=& \tilde\Delta_{n_x}(x)\chi^x\tilde\Delta_{n_y}(y)\chi^y\tilde\Delta_{n_z}(z)\chi^z,
\end{eqnarray}   
\end{subequations}
where functions
\begin{eqnarray}
\tilde \Delta_{n_x}(x) = a\delta(x-(x_{n_x}+a/2)),\label{delta_bra}
\end{eqnarray}
and 
\begin{eqnarray}
\tilde \theta_{n_x}(x) = \left\{\begin{array}{rcl} 1&,&x_{n_x}-a/2<x<x_{n_x}+a/2,\\0&,& \text{otherwise}\end{array}\right..\label{theta_bra}
\end{eqnarray}
Again, functions $\tilde\Delta$ and $\tilde\theta$ for the other two dimensions are defined similarly.

The pairs of functions (\ref{delta_ket}) and (\ref{theta_bra}), and (\ref{theta_ket}) and (\ref{delta_bra}) are such that,
\begin{subequations}
\label{orthogonality_of_theta_Delta}
\begin{eqnarray}
\int_0^{2\pi} dx \Delta_{n_x}(x)\tilde \theta_{n'_x}(x) = \delta_{n_xn'_x}, \\
\int_0^{2\pi} dx \tilde \Delta_{n_x}(x) \theta_{n'_x}(x) = \delta_{n_xn'_x}.
\end{eqnarray}
\end{subequations}

Eqs.(\ref{Bras}) can be thought of as the lattice version of the Hodge conjugation of Eqs.(\ref{Kets}) with respect to the Euclidian metric on the 3-torus. The basis in Eqs.(\ref{Kets}) and (\ref{Bras}) is bi-orthogonal,
\begin{eqnarray}
\langle a,{\bm n}_1|b,{\bm n}_2\rangle = \delta_{ab}\delta^3_{{\bm n}_1{\bm n}_2}.
\end{eqnarray}
Here, the bra-ket overlap is defined as the wedge product of the r.h.s. of Eqs.(\ref{Kets}) and (\ref{Bras}) integrated over the entire $X$. For example,
\begin{eqnarray*}
&\langle 3,{\bm n}|3,{\bm n}\rangle = - \int_X \Delta_{n_x}(x)\chi^x \theta_{n_y}(y)\Delta_{n_z}(z)\chi^z\times\\
&
\times \tilde\theta_{n_x}(x)\tilde\Delta_{n_y}(y)\chi^y\tilde\theta_{n_z}(z) \\
&=\int_0^{2\pi} \Delta_{n_x}(x)\tilde\theta_{n_x}(x)dx\int_0^{2\pi}\theta_{n_y}(y)\tilde\Delta_{n_y}(y)dy\\
&
\times \int_0^{2\pi}\tilde\theta_{n_z}(z)\Delta_{n_z}(z)dz=1,
\end{eqnarray*}
where Eqs.(\ref{orthogonality_of_theta_Delta}) have been used. With this understanding and using the concept of the projection operator,
\begin{eqnarray}
\hat 1_{p} = \sum_{a,{\bm n}}|a,{\bm n}\rangle\langle a,{\bm n}|,\text{  } \hat 1_{p}^2=\hat 1_{p},
\end{eqnarray}
any element of the exterior algebra of $X$ can be projected onto the lattice Hilbert space. For instance, a total probability distribution, $|P\rangle = P({\bm x})\chi^x\chi^y\chi^z$, can be projected onto the sum of the $\delta$-functional distributions,
\begin{eqnarray}
\hat 1_{p} |P\rangle = \sum_{n}P_{\bm n}|1,{\bm n}\rangle,
\end{eqnarray}
with the following self-explanatory coefficients, 
\begin{eqnarray}
P_{\bm n}=\int_{x_{n_x}-a/2}^{x_{n_x}+a/2}dx\int_{y_{n_y}-a/2}^{y_{n_y} + a/2}dy\int_{z_{n_z}-a/2}^{z_{n_z}+a/2}dz P({\bm x}).\nonumber
\end{eqnarray}

The action of the exterior derivative on the basis states can be established straightforwardly,
\begin{eqnarray*}
\hat d |2,{\bm n}\rangle &=& a^{-1}\left(|1,{\bm n}\rangle - |1,{\bm n}+{\bm e}_x\rangle\right),\\
\hat d |3,{\bm n}\rangle &=& a^{-1}\left(|1,{\bm n}\rangle - |1,{\bm n}+{\bm e}_y\rangle\right),\\
\hat d |4,{\bm n}\rangle &=& a^{-1}\left(|1,{\bm n}\rangle - |1,{\bm n}+{\bm e}_z\rangle\right),\\
\hat d |5,{\bm n}\rangle &=& a^{-1}\left(|4,{\bm n}\rangle - |4,{\bm n}+{\bm e}_y\rangle + |3,{\bm n}\rangle - |3,{\bm n}+{\bm e}_z\rangle\right),\\
\hat d |6,{\bm n}\rangle &=& a^{-1}\left(|2,{\bm n}\rangle - |2,{\bm n}+{\bm e}_z\rangle + |4,{\bm n}\rangle - |4,{\bm n}+{\bm e}_x\rangle\right),\\
\hat d |7,{\bm n}\rangle &=& a^{-1}\left(|3,{\bm n}\rangle - |3,{\bm n}+{\bm e}_x\rangle + |2,{\bm n}\rangle - |2,{\bm n}+{\bm e}_y\rangle\right) ,\\
\hat d |8,{\bm n}\rangle &=& a^{-1}\left(|5,{\bm n}\rangle - |5,{\bm n}+{\bm e}_x\rangle \right.\\&&+ \left.|6,{\bm n}\rangle - |6,{\bm n}+{\bm e}_y\rangle+ |7,{\bm n}\rangle - |7,{\bm n}+{\bm e}_z\rangle\right),
\end{eqnarray*}
where ${\bm e}_x=(1,0,0), {\bm e}_y=(0,1,0)$, and ${\bm e}_y=(0,0,1)$. 

As it should, $\hat d$ acts on the Poincar\'e duals of the elements of the cubic CW complex just as the boundary operator would have acted on these elements. Also, the action of $\hat d$ leaves the basis states within the lattice Hilbert space. In other words, one needs no additional projection because 
\begin{eqnarray}
\hat d |i,{\bm n}\rangle =\hat 1_P \hat d \hat 1_P |i,{\bm n}\rangle. \end{eqnarray}
In this sense, $\hat d$ is unique. The actions of all the other operators introduced below, \emph{e.g.}, operator $\chi^x$ defined next, do need the projection onto the lattice Hilbert space. Therefore, all the operators below are essentially the projected operators. For example,
\begin{eqnarray}
\hat \chi^i = \hat 1_P \chi^i \hat 1_P,
\end{eqnarray}
where $\chi^i$ in the r.h.s. denotes the conventional fermionic variable of the exterior algebra of $X$, whereas the l.h.s. denotes the same operator projected onto the lattice Hilbert space.

Using the following properties of the functions introduced above,
\begin{eqnarray}
\int dx \theta_{n_x}(x)\tilde \theta_{n'_x}(x) = \frac12(\delta_{n_xn'_x}+\delta_{n_x(n'_x-1)}),\label{Theta_Tilde_Theta_Overlap}
\end{eqnarray}
it is straightforward to establish the action of operator $\chi^x$ on the basis states,
\begin{eqnarray*}
\hat \chi^x |2,{\bm n}\rangle &=& \left(|1,{\bm n}\rangle+|1,{\bm n}+{\bm e}_x\rangle \right)/2,\\
\hat \chi^x |6,{\bm n}\rangle &=& \left(|4,{\bm n}\rangle+|4,{\bm n}+{\bm e}_x\rangle \right)/2,\\
\hat \chi^x |7,{\bm n}\rangle &=& - \left(|3,{\bm n}\rangle+|3,{\bm n}+{\bm e}_x\rangle \right)/2,\\
\hat \chi^x |8,{\bm n}\rangle &=& \left(|5,{\bm n}\rangle+|5,{\bm n}+{\bm e}_x\rangle \right)/2.
\end{eqnarray*}
The recipe for establishing the above expressions can be demonstrated via the following example,
\begin{eqnarray*}
&\langle 4, {\bm n}'|\chi^x |6,{\bm n}\rangle\\
&=\int_X \tilde\theta_{n'_x}(x)\tilde\theta_{n'_y}(y)\tilde\Delta_{n'_z}(z)\chi^z \chi^x \theta_{n_x}(x)\Delta_{n_y}(y)\chi^y\theta_{n_z}(z)\\
&= \int_0^{2\pi} \tilde\theta_{n'_x}(x)\theta_{n_x}(x)dx\int_0^{2\pi}\tilde\theta_{n'_y}(y)\Delta_{n_y}(y)dy\\&\times \int_0^{2\pi}\tilde\Delta_{n'_z}(z)\theta_{n_z}(z)dz)\\
&=\frac12(\delta_{n_xn'_x}+\delta_{n_x(n'_x-1)})\delta_{n_yn'_y}\delta_{n_zn'_z},
\end{eqnarray*}
as follows from Eqs.(\ref{Theta_Tilde_Theta_Overlap}) and (\ref{orthogonality_of_theta_Delta}).

Similarly, for $\chi^y$ and $\chi^z$ one has,
\begin{eqnarray*}
\hat \chi^y |3,{\bm n}\rangle &=& \left(|1,{\bm n}\rangle+|1,{\bm n}+{\bm e}_y\rangle \right)/2,\\
\hat \chi^y |5,{\bm n}\rangle &=& -\left(|4,{\bm n}\rangle+|4,{\bm n}+{\bm e}_y\rangle \right)/2,\\
\hat \chi^y |7,{\bm n}\rangle &=& \left(|2,{\bm n}\rangle+|2,{\bm n}+{\bm e}_y\rangle \right)/2,\\
\hat \chi^y |8,{\bm n}\rangle &=& \left(|6,{\bm n}\rangle+|6,{\bm n}+{\bm e}_y\rangle \right)/2.
\end{eqnarray*}
and 
\begin{eqnarray*}
\hat \chi^z |4,{\bm n}\rangle &=& \left(|1,{\bm n}\rangle+|1,{\bm n}+{\bm e}_z\rangle \right)/2,\\
\hat \chi^z |5,{\bm n}\rangle &=& \left(|3,{\bm n}\rangle+|3,{\bm n}+{\bm e}_z\rangle \right)/2,\\
\hat \chi^z |6,{\bm n}\rangle &=& -\left(|2,{\bm n}\rangle+|2,{\bm n}+{\bm e}_z\rangle \right)/2,\\
\hat \chi^z |8,{\bm n}\rangle &=& \left(|7,{\bm n}\rangle+|7,{\bm n}+{\bm e}_z\rangle \right)/2.
\end{eqnarray*}

The operator of the flow vector field projected onto the lattice Hilbert space is local in both the spatial and fermionic coordinates,
\begin{eqnarray}
\hat F^i|1, {\bm n}\rangle &=& F^i({\bm x}_{\bm n})|1, {\bm n}\rangle,\\
\hat F^i|2, {\bm n}\rangle &=& F^i({\bm x}_{\bm n}+{\bm e}_x/2)|2, {\bm n}\rangle\\
\hat F^i|3, {\bm n}\rangle &=& F^i({\bm x}_{\bm n}+{\bm e}_y/2)|3, {\bm n}\rangle ,\\
\hat F^i|4, {\bm n}\rangle &=& F^i({\bm x}_{\bm n}+{\bm e}_z/2)|4, {\bm n}\rangle,\\
\hat F^i|5, {\bm n}\rangle &=& F^i({\bm x}_{\bm n}+{\bm e}_y/2+{\bm e}_z/2)|5, {\bm n}\rangle,\\
\hat F^i|6, {\bm n}\rangle &=& F^i({\bm x}_{\bm n}+{\bm e}_x/2+{\bm e}_z/2)|6, {\bm n}\rangle,\\
\hat F^i|7, {\bm n}\rangle &=& F^i({\bm x}_{\bm n}+{\bm e}_y/2+{\bm e}_x/2)|7, {\bm n}\rangle,\\
\hat F^i|8, {\bm n}\rangle &=& F^i({\bm x}_{\bm n}+{\bm e}_x/2+{\bm e}_y/2+{\bm e}_z/2)|8, {\bm n}\rangle,
\end{eqnarray}
where ${\bm e}_x = a(1,0,0), {\bm e}_y = a(0,1,0), {\bm e}_z = a(0,0,1)$, are the spatial shifts on the lattice in the direction.

At this moment, all the necessary operators are defined. The operators can be represented numerically as sparse matrices, and the lattice SEO can then be constructed as,
\begin{eqnarray}
\hat H = [\hat d, \hat F^i \hat \imath_i]_+ + \Theta [\hat d, \hat d^\dagger]_+,\label{SEOLattice}
\end{eqnarray}
where the interior multiplication operator, $\hat \imath_i$, and the co-differential operator, $\hat d^\dagger$, are merely the transpose of the lattice versions of, respectively, $\hat \chi^i$ and $\hat d$. 

Note that the diffusion operator in the lattice SEO (\ref{SEOLattice}) is essentially the Hodge Laplacian. This substitution is not valid in the general case. It is valid, however, for the "Euclidian" vector fields $\bm e$'s in Eq.(\ref{ABC_e_s}).

\end{document}